\documentclass{article}
\def\be{\begin{equation}}
\def\ee{\end{equation}}
\def\m{\mu}
\def\n{\nu}
\def\vta{\vartheta}
\def\a{\alpha}
\def\b{\beta}
\def\c{\gamma}
\def\d{\delta}
\def\Ric{{\rm Ric}}
\def\R{{\rm R}}
\def\nn{\nonumber}
\def\inner{
\hbox{\vbox{\hrule width 4pt } } \! \! \! \!
\hbox{ \vrule height 5pt}\,}
\newcommand{\overstar}{\parbox{0.4ex}{~\raisebox{2.3ex}{$ {}^{\star}$}}
\hspace{0pt}\!\!}
\newcommand{\PR}{\parbox{0.4ex}{~\raisebox{2.5ex}{\tiny${}^{P}$}}
\hspace{0pt}\!\!R\,} 
\newcommand{\LR}{\parbox{0.4ex}{~\raisebox{2.5ex}{\tiny${}^{L}$}}
\hspace{0pt}\!\!R\,}

\begin{document}

\title{The Einstein 3-form $G_\alpha$ and its equivalent 1-form $L_\alpha$ in
Riemann-Cartan space}

\author{Christian Heinicke\\
Institute for Theoretical Physics\\ 
University of Cologne\\ 
50923 K\"oln, Germany\\
Email: chh@thp.uni-koeln.de}

\maketitle

\begin{abstract}
The definition of the Einstein 3-form $G_\alpha$ is motivated by means of
the contracted 2nd Bianchi identity. This definition involves at first the
complete curvature 2-form. The 1-form $L_\alpha$ is defined via 
\mbox{$G_\alpha = L^\beta
\wedge\,^\star(\vartheta_\beta\wedge\vartheta_\alpha)$}
(here $^\star$ is the Hodge-star, $\vta_\a$ the coframe). It is equivalent
to the Einstein 3-form and represents a certain contraction of the curvature
2-form. A variational formula of Salgado on quadratic invariants
of the $L_\alpha$ 1-form is discussed, generalized, and put into proper
perspective.
\end{abstract}

\section{Introduction}

In a Riemannian space, 
the curvature can be split into the conformal (Weyl) curvature and a
piece which contains a 1-form, here called $L_{\a}$. That 1-form
has some very interesting properties: (i) It is closely related to the
Einstein 3-form. (ii) It plays an important role in the
formulation of the initial value problem which focuses on
symmetric-hyperbolic equations for the conformal curvature. (iii)
Recently, we learned of a nice formula of Salgado \cite{Roberto} which
involves an quadratic invariant of $L_{\a}$. 
(iv) It plays a role in the context of current investigations of the 
Cotton 2-form $C_\a$.
The 1-form $L_\a$ appears as the potential of the Cotton 2-form,
$C_\a:=DL_\a$.
The $C_\a$ is 
related to the conformal properties of space and, in 3 dimensions, 
substitutes the Weyl curvature in the criterion for conformal flatness.  

These points motivate a closer look at $L_\a$. We will investigate mainly
its algebraic structure and generalize it to an $n$-dimensional 
Riemann-Cartan space.
The differential properties, that is, the Cotton 
2-form, will be in the center of interest of a forthcoming article
\cite{cottonAA}.

In this article, we would like to cast some light on the Einstein
\mbox{$(n-1)$}-form 
and related quantities. Within the framework of the calculus of exterior
differential forms these structures will arise quite naturally.

In section \ref{bia,ein} we introduce some notation and motivate the
definition of the Einstein
$(n-1)$-form. We give the well known ``differential argument'', involving
the contracted 2nd Bianchi identity, and a less well-known algebraic argument.

In section \ref{alter} we derive two quantities equivalent to the Einstein
3-form, the Einstein tensor and the so-called $L^\a$ 1-form. 

In section \ref{salga} we discuss an invariant containing $L_\a$ and $G_\a$ 
which was
found by Salgado \cite{Roberto} and generalize it to a Riemann-Cartan space. 

Section \ref{irred} puts the previous results into the context of the
irreducible decomposition of the curvature.

We close with a remark concerning the role of the derivative of $L_\a$, 
$DL_\a$, which also is known as Cotton 2-form $C_\a$. 

\section{Bianchi identities and the Einstein $(n-1)$-form}
\label{bia,ein}

On a differentiable manifold of arbitrary dimension we
start with a {\em coframe}\footnote{We use Latin letters
for holonomic and Greek letters for anholonomic indices.}
\be 
\vta ^{\alpha} = {e_i}^{\alpha} \, dx^i \,.
\ee
The coframe is called  natural or holonomic if ${e^{\alpha}}_i =
\delta^{\alpha}_i $. The vector basis or frame which is dual to this
particular coframe is denoted by $e_{\alpha}$,
\be
e_{\alpha}   =   {e^i}_{\alpha} \partial _i \,,\quad
e_{\alpha} \inner \vta ^{\beta}  =  \delta ^{\alpha}_{\beta}\,.
\ee
We then may introduce a {\em connection} 1-form
\be
\Gamma_\a{}^\b = \Gamma_{i\a}{}^\b\,d x^i\,.
\ee
Thereby we define the exterior covariant derivative of a tensor-valued
$p$-form ($d$ denotes the exterior derivative)
\be
D \phi_{\a\,\dots}{}^{\b\,\dots}
:=
d \phi_{\a\,\dots}{}^{\b\,\dots} 
- \Gamma_{\a}{}^\c \wedge \phi_{\c\,\dots}{}^{\b\,\dots} 
+ \Gamma_{\c}{}^\b \wedge \phi_{\a\,\dots}{}^{\c\,\dots}\,.
\ee

Subsequently, we
define the torsion, a vector-valued two-form $T^{\alpha}$ by
\be
T^{\alpha} = \frac{1}{2}\,T_{ij}{}^{\alpha}\,dx^i\wedge dx^j
:= D\vta^{\alpha}= d \vta^{\alpha} +
\Gamma_{\beta}{}^{\alpha}\wedge\vta^{\beta}\,,
\mbox{ $1^{\rm st}$ structure eq.}\,,
\ee
and the curvature, an antisymmetric 2-form ${R_{\alpha}}^{\beta}$ by
\be\label{curv}
{R_{\alpha}}^{\beta} = \frac{1}{2} {R_{ij\alpha}}^{\beta}\, dx^i
 \wedge dx^j := d {\Gamma _{\alpha}}^{\beta} -
 {\Gamma _{\alpha}}^{\gamma} \wedge {\Gamma_{\gamma}} ^{\beta }\,,
\quad\mbox{$2^{\rm nd}$ structure eq.}\,
\ee
From these definitions, together with that of the covariant
exterior derivative, we can deduce the following two identities:
\begin{eqnarray}\label{bia1}
DT^{\alpha}& = & R_{\beta}{}^{\alpha}\wedge\vta^{\beta}\,,
{}\hspace{3.5cm}\mbox{$1^{\rm st}$ Bianchi identity}\,,\\
\label{bia2}
D{R_{\alpha}}^{\beta} & = & 0\,,\hspace{5cm}\mbox{$2^{\rm nd}$ Bianchi
identity}\,.
\end{eqnarray}

Supplied with a {\em metric} $g$ and the corresponding Hodge-star duality 
\mbox{operator ${}^\star$}, we can define the 
$\eta${\em-basis}\footnote{The $\eta${\em-basis} seemingly was introduced by
Trautman, see \cite{Trautman}.}
\scriptsize
\be
\begin{array}{lclcll}
\eta 
& := & 
{}^\star 1\,,
& &
& \mbox{basis of n-forms,}\\
\eta_{\a_1} 
&:=& 
{}^\star \vta_{\a_1} 
&=&
e_{\a_1} \inner \eta\,, 
&\mbox{basis of $(n-1)$-forms,}\\
\eta_{\a_1\a_2} 
&:=& 
{}^\star (\vta_{\a_1}\wedge\vta_{\a_2})
&=&
e_{\a_2}\inner \eta_{\a_1}\,, 
&\mbox{basis of $(n-2)$-forms,}\\
\vdots  &  \vdots  & \vdots & \vdots & \vdots & \qquad\vdots\\
\eta_{\a_1\a_2\dots\a_n} 
&:=& 
{}^\star(\vta_{\a_1}\wedge\vta_{\a_2}\wedge\dots\wedge\vta_{\a_n})
&=&
e_{\a_n}\inner \eta_{\a_1\a_2\dots\a_{(n-1)}}\,.
& 
\end{array}
\ee
\normalsize

\noindent
If we require metric-compatibility of the connection, i.\,e.,
\mbox{$Dg_{\a\b}=0$}, we arrive at a Riemann-Cartan space. In orthonormal
frames, we find \mbox{$\Gamma^{\a\b} = - \Gamma^{\b\a}$}.

By contracting the second Bianchi identity (\ref{bia2}) twice, we find
\be\label{con.bia2}
e_\b \inner e_\a \inner DR^{\a\b} = 0 \quad 
\stackrel{
\begin{array}{c}
\mbox{\footnotesize taking}\\ \mbox{\footnotesize Hodge-dual}
\end{array}
}{\Longrightarrow}
\quad
\vta_\a \wedge \vta_\b \wedge {}^\star DR^{\a\b} =0\,.
\ee
This corresponds to an irreducible piece of the second Bianchi identity, see
\cite{HMCC}. For $n>3$, we obtain another differential identity of the
curvature 2-form by taking the exterior products of 
eq.(\ref{con.bia2}) and 
$\vta_\c$. By using rule H5 for the Hodge-star (see appendix) we find
\be\label{bianchi22}
DR^{\a\b} \wedge \eta_{\a\b\c}=0\,.
\ee 
By differentiation, 
\be
D\left(R^{\b\c}\wedge\eta_{\a\b\c}\right)=
\left(DR^{\b\c}\right)\wedge\eta_{\a\b\c}+R^{\b\c}\wedge
D\eta_{\a\b\c}\,,
\ee 
or, by (\ref{bianchi22}) and $D\eta_{\a\b\c}=T^\d\wedge\eta_{\a\b\c\d}$ (see
\cite{hehl95}, eq.(3.8.5)),
\be\label{dein_rc}
 D\left(\frac{1}{2}\,\eta_{\a\b\c}\wedge R^{\b\c}\right)
= \frac{1}{2}\,\eta_{\a\b\c\d}\,\wedge T^\b \wedge R^{\c\d}\,.
\ee 
In a Riemann space,  where the torsion is zero, {\em and} in a Weitzenb\"ock 
space, where the curvature is zero, the term on the right hand side vanishes. 

Another interesting property of the $(n-1)$-form 
$\eta_{\a\b\c}\wedge R^{\b\c}$ should be mentioned here. 
From special relativity we know that the energy-momentum current density has
to be represented by a vector-valued $(n-1)$-form. Suppose one has the 
idea to link  energy-momentum  to curvature.
We then notice that
the expression $\eta_{\a\b\c}\wedge R^{\b\c}$ is one of the two 
most obvious vector valued $(n-1)$-forms linear in the curvature.
The other one is $\vta_\b\wedge{}^\star R^{\a\b}$.
However, in a Riemannian space, only the first 
quantity is automatically conserved\footnote{
In a {\em Riemannian} space we find for  the covariant derivative of
$\vta_\b \wedge {}^\star R^{\a\b}$, using the 
notation introduced in  section \ref{alter}, 
$D\left(\vta_\b\wedge{}^\star R^{\a\b}\right) 
=-D{}^\star\!\left(e_\b \inner R^{\a\b} \right)
=-D{}^ \star\!\left(\Ric_\nu{}^\a\,\vta^\nu\right) 
=-(\nabla _\mu \Ric_\nu{}^\a) \,\vta^\mu\wedge\eta^\nu 
=-(\nabla _\mu \Ric_\nu{}^\a) \,g^{\nu\mu}\,\eta 
=-\frac{1}{2}\,(\nabla^\a \R) \, \eta $.
The last expression, which is non-zero in general, 
is obtained by means of the 2$^{\rm nd}$ Bianchi identity.}
which perfectly matches the conserved energy-momentum.\footnote{
In a Riemann-Cartan space appear additional forces on the right-hand
side of the energy-momentum law, such as the {\em Mathisson-Papapetrou force}. 
This is consistent with the non-vanishing right-hand side of
eq.(\ref{dein_rc}), see \cite{HMCC}.}

These considerations motivate the definition\\

\framebox{\parbox{10cm}{
\be\label{ein}
G_{\a} := \frac{1}{2}\,\eta_{\a\b\c}\wedge R^{\b\c}\,,
\qquad\mbox{\em Einstein $(n-1)$-form.} 
\ee}}

\section{Alternative representations of the Einstein\\
\mbox{\boldmath$(n-1)$}\unboldmath-form}
\label{alter}

The $(n-1)$-form (\ref{ein}) naturally appears as a piece of the identically
vanishing \mbox{$(n+1)$-form} $R^{\b\c}\wedge\eta_{\a}$. 
We try to extract $G_\a$ from $R^{\b\c}\wedge\eta_{\a}$
by contracting the latter twice:
\begin{eqnarray}\label{einl}
0
&=&
e_\c\inner e_\b\inner\left( R^{\b\c}\wedge\eta_{\a}\right) 
=e_\c \inner \left[   \left( e_\b \inner R^{\b\c}\right)\wedge\eta_{\a}
                   +  R^{\b\c} \wedge \eta_{\a\b}
            \right] \nn\\
&=&
\underbrace{\left( e_\c \inner e_\b \inner R^{\b\c} \right)
            }_{=-\R}
 \wedge \eta_\a
- \underbrace{\left( e_\b \inner R^{\b\c}\right)}_{=-\Ric^\c}
   \wedge \eta_{\a\c}
+ \underbrace{\left( e_\c \inner R^{\b\c}\right)}_{=\Ric^\b}
   \wedge \eta_{\a\b}
+ \underbrace{R^{\b\c} \wedge \eta_{\a\b\c}}_{=2G_\a}\,,\nn\\
\end{eqnarray}
\be\label{einl2}
\Rightarrow\,\, G_\a  =  -\Ric^{\b} \wedge \eta_{\a\b}
                           + \frac{1}{2} \, \R \, \eta_{\a}\,,
\ee
where we  introduced the {\em Ricci 1-form} 
$\Ric_\a:=e_\b \inner R_\a{}^\b\,$, and its trace, the {\em curvature
scalar}\/ $\R := e_\a \inner \Ric^\a$\,.
In a Riemann-Cartan space we have \mbox{$R^{\a\b}=-R^{\b\a}$}.
By means of the definition of the curvature 2-form, eq.(\ref{curv}), we find
\mbox{$\Ric_{\a} = \Ric_{\n\a}\,\vta^{\n}$}, where \mbox{$\Ric_{\n\a} :=
R_{\m\n\a}{}^{\m}$} denotes the Ricci tensor.
It is symmetric if the torsion is covariantly constant as, for instance, 
in a Riemann space, where $T^\a\equiv 0$.\\

Often the Ricci tensor is defined by contraction of the 2nd and 4th index 
in our Schouten notation of
the curvature tensor. Because of the antisymmetry of the curvature tensor, 
this definition of the Ricci tensor differs from our convention by a sign.
This applies also to quantities which are derived from the Ricci tensor, like
the Einstein tensor and the $L_{\a\b}$ tensor.

\subsection*{Einstein tensor}
Since $G_\a$ is a $(n-1)$-form, we can decompose it with respect to the basis of
$(n-1)$-forms $\eta_\a$:
\be\label{def:einten}
G_\a =: G_\a{}^\b\,\eta_\b\,.
\ee
By convention, we contract here the 2nd index of $G_\a{}^\b$.
In order to determine the components $G_\a{}^\b$,
we just have to rewrite the first term of
the right hand side of eq.(\ref{einl}):
\begin{eqnarray}\label{riceta}
\Ric^\b \wedge \eta_{\a\b} 
&=&
 \Ric_\m{}^\b\,\vta^\m \wedge \eta_{\a\b} =
\Ric_\m{}^\b\,\vta^\m \wedge {}^{\star} (\vta_\a \wedge \vta_\b)\nn\\
&=& 
- \Ric_\m{}^\b\, {}^{\star} \left(e^\m \inner (\vta_\a \wedge\vta_\b)\right)
= - \Ric_\m{}^\b\, {}^{\star} (\d_\a^\m\,\vta_\b - \vta_\a \,\d_\b^\m)\nn\\
&=&
 - \Ric_\a{}^\b\,\eta_\b + \R \, \eta_\a\,.
\end{eqnarray}
By substituting (\ref{riceta}) into (\ref{einl}) we find
\be\label{einten}
G_{\a\b} = \Ric_{\a\b} - \frac{1}{2}\,\R \,g_{\a\b}\,.
\ee
Thereby we recover the usual definition of the Einstein tensor.

\subsection*{The $L_\a$ 1-form}
Using the identity $ \vta^{\a} \wedge (e_\a \inner \Phi) = 
\mbox{\scriptsize$({\rm rank\,}\Phi)$\normalsize}\,\,
 \Phi$, we can rewrite $\eta_\a$ according to
\be\label{theta_wedge_eta}
(n-1)\, \eta_\a = \vta^{\b}\wedge (e_\b \inner \eta_\a) = \vta^{\b} \wedge
\eta_{\a\b}\,.
\ee
Substituting this into (\ref{einl2}), we arrive at
\be
G_\a =  \Ric^{\b}\wedge\eta_{\b\a} -
\frac{1}{2(n-1)}\,\R\,\vta^\b\wedge\eta_{\b\a}\,.
\ee

This suggests the definition\\

\framebox{\parbox{10cm}{\medskip
\be
L^\b :=  \Ric^{\b} - \frac{1}{2(n-1)}\,\R\,\vta^\b\,,
\ee}}

\medskip
such that
\be\label{gofl}
G_\a = L^{\b} \wedge \eta_{\b\a}\,.
\ee

\medskip

\noindent
The decomposition of $L_\a$ in components reads
\be
L^\a =: L_\m{}^\a\,\vta^\m\,.
\ee
In a Riemann-Cartan space, the tensor
\be\label{lten}
L_\m{}^\a = \Ric_\m{}^\a -
\frac{1}{2(n-1)} \,\R \d_{\m}^\a\,,
\ee
if $\alpha$ is lowered, 
is {\em not} symmetric in general. 

The trace of $L_\a$ is proportional to the curvature scalar
\be
L:= e_\a \inner L^\a = \frac{n-2}{2(n-1)}\,\R\,.
\ee

We collect the results of this section for a Riemann-Cartan space in

\begin{center}
\framebox{\parbox{6cm}{
\be\label{ein,r,l,g}
\begin{array}{rcccl}
G_\a &=& 
\frac{1}{2}\,R^{\b\c} 
&\wedge&
\eta_{\a\b\c}\vspace{3pt}\\
&=& 
L^\b 
&\wedge& 
\eta_{\b\a}\vspace{3pt}\\
&=&
G_\a{}^\b 
&\wedge& 
\eta_\b\,.
\end{array}
\ee}}
\end{center}
\section{On Salgado's formula}
\label{salga}

Since $G_\a$ is a $(n-1)$-form and $L^\a$ a 1-form, $L^\a\wedge G_\a$ is
a scalar-valued
$n$-form and thus a possible candidate for a {\em Lagrange n-form}.
Moreover, we can guess that the variation of $L^\a\wedge G_\a$ with respect
to $L^\a$ should yield $G_\a$, and vice versa. However, because $L_\a$ and
$G_\a$ are not independent, we have to check this explicitly.
By means of the results of the last section we find
\begin{eqnarray}
\d\,\frac{1}{2}\left( L^\a\wedge G_\a\right)
&=&
\frac{1}{2}\left(
\d L^\a \wedge G_\a + L^\a \wedge \d G_\a
\right)\nn\\
&=&
\frac{1}{2}\left(
\d L^\a \wedge G_\a + L^\a \wedge \d\left(L^\c \wedge
\eta_{\c\a}\right)
\right)\nn\\
&=&
\frac{1}{2}\left(
\d L^\a \wedge G_\a + L^\a \wedge \d L^\c \wedge
\eta_{\c\a} + L^\a \wedge L^\c \wedge \d \eta_{\c\a}
\right)\nn\\
&=&
\frac{1}{2}(
\d L^\a \wedge G_\a - \d L^\c \wedge 
\underbrace{L^\a\wedge\eta_{\c\a}}_{=-G_\c}
+L^\a \wedge L^\c \wedge \d \eta_{\c\a})\nn\\
&=&
\d L^\a \wedge G_\a - \frac{1}{2} \,L^\a\wedge L^\b\wedge\d\eta_{\a\b}\,.
\end{eqnarray}
Thus we have\\

\begin{center}
\framebox{\parbox{7cm}{\vspace{0.3cm}
\be\label{vardev}
\frac{1}{2}\,\frac{\delta\,(L^\a\wedge G_\a)}{{\rm\delta}L^\b} = G_\b\,.
\ee}
}
\end{center}
  
\noindent
This formula also becomes apparent by noticing that 
\be 
\frac{1}{2}\,L^\a \wedge G_\a =- \frac{1}{2}\, L^\a \wedge L^\b \wedge
\eta_{\a\b}\,.
\ee

For the variation of $L^\a\wedge G_\a$ with respect to $G_\a$ we have to
express $L^\a$ in terms of $G_\a$. We start from
\be 
G_\a = L^\b\wedge \eta_{\b\a} = L_\m{}^\b\,\vta^\m \wedge \eta_{\b\a}\,,
\ee
The term $\vta^\m \wedge \eta_{\b\a}$ can be rewritten as in eq.(\ref{riceta})
yielding
\be \label{leineta}
G_\a = L_\a{}^\b\, \eta_\b - L\,\eta_\a
=\Big(L_\a{}^\b - L \d_\a^\b\Big)\,\eta_\b\,.
\ee
From this equation we infer for the traces $L$ and $\overstar G$ 
\be \label{gltrace}
\overstar G := e_\a \inner {}^\star G^\a={}^\star(G^\a\wedge\vta_\a) 
= (-1)^{(n-1+{\rm ind})}\,\,(1-n)\,L\,.
\ee
The last two equations lead to
\begin{eqnarray}\label{lgtensor}
L_\a{}^\m &=& L_\a{}^\b\,\d_\b^\m = L_\a{}^\b \, e^\m \inner \vta_\b
= L_\a{}^\b \, e^\m \inner \, \left((-1)^{(n-1+{\rm ind})}\, \,{}^\star 
\eta_\b\right) \nn\\
&=&
(-1)^{(n-1+{\rm ind})}\,\, \Big( e^\m \inner {}^\star G_\a -\frac{1}{n-1}\,
\overstar G\,\d_\a^\m \Big)\,,
\end{eqnarray}
or, by multiplying with $\vta^\a$ and using the rules for the
Hodge-dual,\footnote{If $DT^\a\equiv 0$, $L_{\a\b}$ is symmetric and we
simply have\\ ${}^\star L^\a = {}^\star (L_\m{}^\a\,\vta^\m)
= L_\m{}^\a \, {}^\star \vta^\m
= L_\m{}^\a \,\eta^\m=L^\a{}_\m\,\eta^\m\,,$ or, by substituting this into 
eq.(\ref{leineta}), $G_\a = {}^\star L_\a - L \eta_\a\,.$
}
\be\label{lofg}
L^\m = (-1)^{{\rm ind}}\,\,{}^\star\Big[e^\a \inner (G_\a\wedge\vta^\m) 
- \frac{1}{n-1} \, e^\m \inner (G_\a \wedge \vta^\a)\Big]\,.
\ee
Since Hodge-star, interior and exterior products are linear, eq.(\ref{lofg})
is linear in $G_\a$. Consequently, the variation of $L_\a$ with 
respect to $G_\a$ reads
\be\label{lvarg}
\d L^\a (G_\b) = L^\a(G_\b + \d G_\b) - L^\a (G_\b) 
              =  L^\a(\d G_\b)\,.
\ee
A simple, but somewhat lengthy, calculation shows\footnote{See appendix.}
\be 
L^\a(\d G_\b) \wedge G_\a = L^\a \wedge \d G_\a \,.
\ee
Then the variation of $L^\a \wedge G_\a$ with respect to $G_\a$ turns out to
be
\be
\frac{1}{2}\,\d \, (L^\a\wedge G_\a) = 
\frac{1}{2}\,\left[ (\d L^\a) \wedge G_\a + 
L^\a \wedge (\d \,G_\a) \right]
= L^\a \wedge \d G_\a \,.
\ee

\begin{center}
\framebox{\parbox{7cm}{\medskip
\be\label{vardev2}
\frac{1}{2}\,\frac{\delta\,(L^\a\wedge G_\a)}{{\rm\delta}G_\b} = 
(-1)^{(n-1)}\,L^\b\,.
\ee\medskip}}
\end{center}

\subsection*{Component representation}
Evaluating $e_\b \inner e_\a \inner (L^\a \wedge L^\b \wedge \eta)=0$ yields
\begin{eqnarray}\label{sal0}
  \frac{1}{2}\, L^\a \wedge G_\a 
&=& -\frac{1}{2}\, L^\a \wedge L^\b \wedge \eta_{\a\b} 
= -\frac{1}{2}\left( L^{\a}{}_{\a}\,L^{\b}{}_{\b}
                    - L^{\a}{}_{\b}\,L^{\b}{}_{\a}
              \right)\eta\,\nn\\
&=& -\left(L^{\a}{}_{[\a}\,L^{\b}{}_{\b]}\right)\eta\,,
\end{eqnarray}
Eq.(\ref{vardev}) corresponds to the Salgado formula \cite{Roberto} 
\be\label{salgado}
\frac{d\left(L^{\a}{}_{[\a}\,L^{\b}{}_{\b]}
    \right)}{dL^{\m}{}_\n}=-L^\n{}_\m+\delta_\m^\n\,L= -G^\n{}_\m\,,
\ee
which was found by Salgado in a Riemannian context. It remains 
valid in a Riemann-Cartan space. 
Eqs.(\ref{einten}, \ref{lten}) yield
\be\label{l=ein}
L^\a{}_\b = G^\a{}_\b -\frac{1}{n-1}\,G^\m{}_\m\,\d^\a_\b\,.
\ee
Differentiating the last equation we get
\be
\frac{\partial\, L^\a{}_\b}{\partial\,G^\m{}_\n} = \d^\a_\m\,\d_\b^\n
-\frac{1}{n-1}\d^\n_\m\,\d^\a_\b\,.
\ee
Substituting eq.(\ref{l=ein}) into $L^{\a}{}_{[\a}\,L^{\b}{}_{\b]}$\,, we find
\be
L^{\a}{}_{[\a}\,L^{\b}{}_{\b]} = -\frac{1}{2} L^{\a}{}_{\b}\,G^{\b}{}_{\a}\,.
\ee
From the last two equation we derive 
\be
\frac{d}{d G^\m{}_\n} \left(L^{\a}{}_{[\a}\,L^{\b}{}_{\b]}\right) =
-L^{\n}{}_{\m}\,.
\ee

\section{The 1-form $L_\a$ and the irreducible decomposition of the curvature}
\label{irred}

The 1-form $L_\a$ represents the trace-part (that is, the second rank pieces of a
fourth rank quantity) 
$e_\b \inner R^{\a\b} = L^\a + \frac{1}{n-2}L \vta^\a$ of the curvature. 
This property seems
to be nothing special because it also belongs to other contractions of the
curvature (like the Ricci- and the Einstein-tensor). However, the Einstein
tensor, which is a trace-modified Ricci-tensor, is an
interesting quantity because of a property not shared by
the Ricci-tensor, namely to be divergence-free.
What are the properties peculiar to $L_\a$? 

In a Riemann-Cartan space, 
the 1-form $L_\a$ represents that part of the curvature 2-form which has the
structure $\vta_{[\a}\wedge({\rm1-form})_{\b]}$. To see this, one has to
perform an irreducible decomposition of the curvature. We use the results
obtained in \cite{McCrea} and find\\

\small
\noindent
\framebox{\parbox{12cm}{
\be\label{decomp}
\begin{array}{ccccccc}
\R_{\a\b} 
&=& 
{}^{(1)}R_{\a\b} 
&+&
{}^{(2)}R_{\a\b}+{}^{(3)}R_{\a\b}
&+&
{}^{(4)}R_{\a\b} + {}^{(5)}R_{\a\b} + {}^{(6)}R_{\a\b}$\vspace{5pt}$\\
&=&                                                              
{}^{(1)}R_{\a\b}       
&+&
(-1)^{\rm\tiny ind}\,
{}^{\star}\left(\vta_{[\a}\wedge P_{\b]}\right)
&-&
\frac{2}{n-2} \vta_{[\a}\wedge L_{\b]}$\vspace{5pt}$\\
&=:&
{\rm Weyl}_{\a\b}
&+&
\PR_{\a\b}
&+&
\LR_{\a\b}$\vspace{5pt}$\\
&=&
\begin{array}{l}
\mbox{irreducible}\\ 
\mbox{$4^{\rm\tiny th}$-rank}
\end{array}
&+&
\mbox{pseudo-trace piece}
&+&
\mbox{trace piece}
\end{array}
\ee}}
\normalsize

\medskip
The ${}^{(i)}R_{\a\b}$ denote the 6 irreducible pieces of the curvature in a
Riemann-Cartan space. For their precise definition we refer the reader to
the literature, see \cite{McCrea,hehl95}, e.\,g., because, in this context, the
main result is contained in the second line of (\ref{decomp}).  
The curvature decomposes into the conformal curvature $\rm Weyl_{\a\b}$, 
a trace piece piece $\LR_{\a\b}$, determined by $L_\a$, and a pseudo-trace
piece $\PR_{\a\b}$, which is determined by a $(n-3)$-form\footnote{For
the sake of completeness we display its explicit form:\\ 
$P_\a := {}^\star (R^{\b}{}_\a\wedge\vta_\b)
-\frac{1}{6}\,\vta_\a\wedge{}^\star(R^{\b\c}\wedge\vta_\b\wedge\vta_\c) -
\frac{1}{n-2}\,e_\a \inner
\left[\vta^\b\wedge{}^\star(R^\c{}_\b\wedge\vta_\c)\right]$}
$P_\a$.
If $DT^\a \equiv 0$, i.\,e., in particular in a Riemannian space,
we have, due to the 1st Bianchi identity (\ref{bia1}),
${}^{(2)}R_{\a\b}={}^{(3)}R_{\a\b}={}^{(5)}R_{\a\b}=0$ and thus
\be\label{r=w+l}
R_{\a\b} = {\rm Weyl}_{\a\b} - \frac{2}{n-2}\,\vta_{[\a}\wedge L_{\b]}\,,
\qquad\mbox{if $T^\a\equiv0$.}
\ee
The corresponding formula in Ricci calculus is often used for
defining the Weyl tensor $C_{\a\b\c\d}$. Eq.(\ref{r=w+l}),
decomposed into components reads
\be
C_{\a\b\c\d}
=
R_{\a\b\c\d} +
\frac{4}{n-2}\,g_{\big[\a\big|[\c}L_{\d]\big|\b\big]}\,,
\qquad\mbox{if $T_{\a\b}{}^\c\equiv0$.}
\ee

In a Riemannian space, ${\rm Weyl}_{\a\b}$ transforms like a conformal
density. Thus, $L_\a$ represents that piece of the 
curvature which does not transform like a conformal density.
This properties of $L_\a$ are well known, compare \cite{Eisenhart} 
and \cite{Schouten}.

For the number of independent components we have
\begin{eqnarray}
{\rm Weyl}_{\a\b} & \to & \frac{1}{12}(n+2)(n+1)n(n-3)\,,\\
\PR_{\a\b} & \to & \frac{1}{6}\,(n+1)(n-1)n(n-3)\,,\\
\LR_{\a\b} & \to & n^2\,.
\end{eqnarray}
These pieces have characteristic trace properties
\be
e_\a \inner {\rm Weyl}^{\a\b}
=
e_\a \inner \PR^{\a\b} = 0\,,\quad 
e_\a \inner \LR^{\a\b} = e_\a \inner R^{\a\b}\,.
\ee
By means of those we find ${\rm Weyl}^{\b\c}\wedge\eta_{\a\b\c}= 
\PR^{\b\c}\wedge\eta_{\a\b\c}=0$.
Thus, only the piece $\LR^{\a\b}$ contributes to the Einstein 3-form.
This is even more apparent by substituting $\vta^\c\wedge\eta_{\a\b\c} =
(n-2)\,\eta_{\a\b}$ into eq.(\ref{ein,r,l,g}),   
\begin{eqnarray}\label{g(lr)}
G_\a
&=& 
L^\b\wedge\eta_{\b\a}
=
\frac{1}{n-2}\,L^\b\wedge\vta^\c\wedge\eta_{\b\a\c}
=  
-\frac{1}{n-2}\,\vta^{[\b}\wedge L^{\c]}\wedge\eta_{\a\b\c}\,,\nn
\end{eqnarray}
or, by using (\ref{decomp}),
\begin{center}
\framebox{\parbox{6cm}{\medskip
\be
G_\a = \frac{1}{2}\,\LR^{\b\c}\wedge\eta_{\a\b\c}\,.
\ee
}}
\end{center}

We can use this relation in order to obtain another well motivated
representation of the invariant $L^\a\wedge G_\a$ by rewriting it according
to\footnote{We use $(n-3)\,\eta_{\a\b\c} =\vta^\d \wedge
\eta_{\a\b\c\d}$\,.}
\begin{eqnarray}
I_S &:=& -L^{\a}{}_{[\a}L^{\b}{}_{\b]}\,\eta= 
\frac{1}{2}\, L^\a \wedge G_\a
= \frac{1}{4}\, L^\a \wedge \eta_{\a\b\c}\wedge \LR^{\b\c}\nn\\
&=&\label{lr**2}   
-\frac{n-2}{8(n-3)} \, \LR^{\a\b}\wedge \LR^{\c\d}\,\eta_{\a\b\c\d}\,.
\end{eqnarray}     
In this way, $I_S$ turned out to be one of the basic quadratic 
invariants\footnote{In the case
$n=4$ we can define the {\em Lie-dual}\/ of \/
\mbox{$\LR_{\a\b}$ by $\LR_{\a\b}^\ast
:= \frac{1}{2}\,\LR^{\c\d}\,\eta_{\a\b\c\d}$}. Then $I_S$ reads
$I_S=-\frac{1}{2}\,\LR^{\a\b}\wedge \LR_{\a\b}^\ast\,.$
Using the irreducible decomposition (\ref{decomp}), $I_S$ can also be
expressed in terms of the Hodge-dual
\mbox{$I_S \propto
R^{\a\b}\wedge\,{}^\star\left({}^{(4)}R_{\a\b}-{}^{(5)}R_{\a\b}
-{}^{(6)}R_{\a\b}\right)\,.$}
If $DT^{\a}\equiv0$, we can use $G_\a={}^\star L_\a -L\,\eta_\a$ and obtain
$I_S= \frac{1}{2}\,[L^\a\wedge {}^\star L_\a - (L)^2\,\eta]=
\frac{1}{2}\left[L_{\nu}{}^{\alpha}\,L^{\nu}{}_{\alpha} - (L)^2
\right]\,\eta\,.$
The positions of the indices differ from those in eq.(\ref{sal0})!
}
(scalar-valued $n$-forms) of $\LR_{\a\b}$.

\noindent
We collect the various representation of the invariant $I_S$ in

\medskip

\noindent
\framebox{\parbox{11.5cm}{
\be\label{I_S}
I_S = \frac{1}{2}\,L^\a\wedge G_\a = - \frac{1}{2}\, L^\a\wedge L^\b \wedge
\eta_{\a\b} = - \frac{n-2}{8(n-3)} \, \LR^{\a\b}\wedge
\LR^{\c\d}\,\eta_{\a\b\c\d}\,.
\ee}}

\section{Discussion}

In view of our observations, we may put the main result of our
investigations as follows. The basic quantity here is the ``trace-part''
$\LR_{\a\b}$ of the curvature  with
its $n^2$ independent components. A vector-valued 1-form, a vector-valued 
$(n-1)$-form, and a 2nd rank tensor valued 0-form, respectively,
have the same number of independent components. Thus $\LR_{\a\b}$, as
displayed in eq.(\ref{g(lr)}) and in eq.(\ref{ein,r,l,g}), can 
be mapped into a $(n-1)$-form by means of the $\eta$-basis, yielding the
Einstein 3-form $G_\a$, into a 1-form, yielding the $L_\a$ 1-form, and into
an $n\times n$ matrix, yielding the Einstein tensor $G_{\a\b}$.
Realizing this, makes the algebraic relations between the stated quantities
quite clear. These results are represented by Eqs.(\ref{ein,r,l,g},
\ref{g(lr)}), and (\ref{decomp}). 

We also would like to mention that eq.(\ref{gofl}) which expresses $G_\a$ in
terms of $L_\b$ and eq.(\ref{lofg}) which expresses $L_\a$ in terms of
$G_\a$ hold for arbitrary $(n-1)$ forms $G_\a$ and 1-forms $L_\b$,
respectively. Hence, eq.(\ref{gofl}) and eq.(\ref{lofg}) establish a 
{\em duality relation} between $(n-1)$-forms and 1-forms.

The invariant $L^\a{}_{[\a}\,L^{\b}{}_{\b]}$, which was derived by
Salgado as second principal invariant of $L^\a{}_\b$ (arising in connection
with the characteristic polynomial), in our context 
emerges (i) as one of the
the most obvious invariants constructed from $L_\a$ and $G_\a$, (ii) as a
basic quadratic invariant of $L_\a$, and (iii) as a
basic quadratic invariant of the curvature piece $\LR_{\a\b}$.
These results are displayed in (\ref{I_S}) and (\ref{vardev}, \ref{vardev2}).

We have extensively studied the algebraic properties of $L_\a$.
It was quite helpful to check all formulas by means of the Excalc package of
the computer algebra system Reduce, see \cite{smh}.

Also the differential properties of $L_\a$ are very interesting. 
In a Riemannian space we have $D\vta^{\a}=0$. By using (\ref{r=w+l}),
we can represent the second Bianchi Identity as follows
\be
D{\rm Weyl}_{\a\b} = -\frac{2}{n-2}\,\vta_{[\a}\wedge C_{\b]}\,,
\ee
where we defined the Cotton 2-form by
\be\label{def:cotton}
C_\a :=  D L_\a\,.
\ee
In this way, $L_\a$ appears as {\em potential} of the Cotton
2-form. 
Since the conformal (Weyl-) curvature is tracefree, the (twice) contracted
2nd Bianchi identity reads
\be\label{cottontrace}
0=e_\a \inner C^\a = e_\a \inner DL^\a\,.
\ee
The Cotton 2-form, especially its relation to the
conformal properties of spacetime, is subject of a current project
\cite{cottonAA}. The definition (\ref{def:cotton}) of the Cotton 2-form can
be transferred to a Riemann-Cartan space.

\bigskip

\noindent
{\bf Acknowledgements}

   The author is grateful to F.\,W.~Hehl for many helpful discussions. 
The author would also like to thank G.\,F.~Rubilar and
Y.\,N.~Obukhov for useful remarks.

\section{Appendix}

\subsection{Some relations for the exterior and interior products}

In order to avoid dimension-dependent signs, it is of special importance to
take care of the order of the  forms in the exterior products.
We would like to remind the reader of the following relations which hold for
a $p$-form $\phi$ and a $q$-form $\psi$:

\be
\phi \wedge \psi = (-1)^{pq}\,\psi \wedge \phi\,,
\ee
\be
e_\m \inner (\phi\wedge\psi) = (e_\m \inner \phi ) \wedge \psi
 + (-1)^p\, \phi \wedge (e_\m \inner \psi)\,.
\ee

\subsection{The variational derivative with respect to p-forms} 

The variation of a function $F$ which depends on a $p$-form $\psi$ is
defined to be
\be\label{def:vardef}
\d F := F(\psi+\d \psi) - F(\psi)\,,
\ee
where the $p$-form $\d \psi$ is supposed to be an arbitrary ``small'' 
deviation. With given $F$, we can elementary evaluate the right-hand side of
eq.(\ref{def:vardef}). We then neglect all terms of quadratic and higher order
in $\d \psi$ and bring the result into the form
\be
\d F = \d\psi \wedge (\, \dots \,)\,.
\ee
The expression in the parentheses is defined to be the partial 
derivative with respect to $\psi$. This prescription especially fixes the
sign. The generalization to an arbitrary number of forms or tensor-valued
forms is straightforward. 

Due to the definition, the variation obeys a Leibniz-rule
\begin{eqnarray}
\d (\phi\wedge\psi) 
&\stackrel{(\ref{def:vardef})}{=}&
\left( (\phi + \d \phi ) \wedge ( \psi\ + \d \psi ) \right)
- \phi\wedge\psi\nn\\
&=& 
\phi \wedge \d \psi + \d\phi \wedge \psi + \underbrace{\d\phi
\wedge\d\psi}_{\hookrightarrow 0}\nn\\ 
&=& 
\phi \wedge \d \psi + \d\phi \wedge \psi\,.
\end{eqnarray}
 
The variational derivative can be introduced in the usual way. However,
in this context we just note that in the case in which $F$ does not depend on
the derivatives $d\psi$, the partial and the variational derivative coincide.

\subsection{Some relations for the Hodge-star}

We frequently made use of the following relations for the Hodge-star.
$\psi$ an $\phi$ are two p-forms of the same degree, $a,\,b \in {\bf R}$ 
are numbers.

\be
{}^\star(a\psi + b\phi) = a {}^\star \psi + b {}^\star \phi\,,
\qquad\mbox{H1}\,.
\ee

\be\label{h2}
{}^\star {}^\star \psi = (-1) ^{p(n-p)+{\rm {\rm ind}}}\,
\psi\,,\qquad\mbox{H2}\,,
\ee\small
   where {\scriptsize {\rm ind}} denotes the number of negative 
Eigenvalues of  the metric, 3 in the case of a $(3+1)$-dimensional spacetime.
\normalsize

\be\label{h3}
{}^\star (e_\a \inner \psi) = (-1) ^{(p-1)}\, \vta_a \wedge {}^\star
\psi\,,\qquad\mbox{H3\,.} 
\ee

\be\label{h4}
e_\a \inner{}^\star \psi =  {}^\star (\psi \wedge \vta_\a)\,,
\qquad \mbox{H4}\,.
\ee

\be\label{h5}
{}^\star \psi \wedge \phi = {}^\star \phi \wedge \psi\,,\qquad\mbox{H5}\,.
\ee

\subsection{Variation of \boldmath $L_\a$}

We substitute eq.(\ref{lofg}) in eq.(\ref{lvarg}):
\be
\d L^\a =  L^\a(\d G_\b) = 
(-1)^{{\rm ind}}\,\,{}^\star\!\left[ e^\b\inner(\d G_\b\wedge \vta^\a) -\frac{1}{n-1}\,e^\a \inner (\d
G_\b \wedge \vta^\b)\right] \,.
\ee
The expression in the square-brackets is a $(n-1)$-form. By H5 we then have
\begin{eqnarray}\label{lg1}
\d L^\a \wedge G_\a 
&=& 
(-1)^{{\rm ind}}\,\,{}^\star\!\left[ e^\b\inner(\d G_\b\wedge \vta^\a) 
-\frac{1}{n-1}\,e^\a \inner (\d G_\b \wedge \vta^\b)\right] 
\wedge G_\a\nn\\
&=&
(-1)^{{\rm ind}}\,\,\left\{{}^{\star} G_\a \wedge \Big[e^\b \inner 
(\d G_\b\wedge\vta^\a)\Big] - \frac{1}{n-1}\,
{}^{\star} G_\a \wedge \Big[e^\a \inner (\d G_\b\wedge\vta^\b)\Big]
\right\}\nn\,.\\
&&
\end{eqnarray}
${}^\star G_\a$ is a 1-form and $\d G_\b\wedge \vta^\b$, 
$\d G_\b\wedge \vta^\b$ are $n$-forms. Thus
\begin{eqnarray}
0 
&=&
e^\b \inner \Big[{}^{\star} G_\a \wedge(\d G_\b\wedge\vta^\a)\Big]
= (e^\b \inner {}^{\star} G_\a)\, \d G_\b \wedge \vta^\a 
- {}^{\star} G_\a \wedge \Big[e^\b \inner (\d G_\b \wedge
\vta^\a)\Big]\nn\,,\\
0
&=&
e^\a \inner \Big[{}^{\star} G_\a \wedge(\d G_\b\wedge\vta^\b)\Big]
= (e^\a \inner {}^{\star} G_\a)\,\d G_\b \wedge \vta^\b
- {}^{\star} G_\a \wedge \Big[e^\a \inner (\d G_\b\wedge \vta^\b)\Big]\nn\,.
\end{eqnarray} 
Substituting this into eq.(\ref{lg1}) yields
\begin{eqnarray}
\d L^\a \wedge G_\a 
&=&
(-1)^{{\rm ind}}\,\,(-1)^{(n-1)} \, 
\Big[ (e^\b \inner {}^\star G_\a)\, \vta^\a 
-\frac{1}{n-1}\,(e^\a \inner {}^\star G_\a)\,\vta^\b\Big] \wedge \d
G_\b\nn\\
&=&
(-1)^{{\rm ind}}\,\,(-1)^{(n-1)} \,\Big[  {}^{\star} 
( G_\a \wedge \vta^\b ) \, \vta^\a 
-\frac{1}{n-1}\,{}^\star (G_\a\wedge \vta^\a)\, \vta^\b \Big]
\wedge \d G_\b\nn\\
&=&
(-1)^{{\rm ind}}\,\,{}^\star\!\Big[ e^\a \inner (G_\a\wedge\vta^\b) 
- \frac{1}{n-1}\, e^\b \inner 
(G_\a \wedge \vta^\a)\Big] \wedge \d G_\b\nn\\
&=&
L^\b \wedge \d G_\b\nn\\
&=&
(-1)^{(n-1)}\, \d G_\b \wedge L^\b\,.
\end{eqnarray}

\end{document}